# Observation of Square-Root Higher-Order Topological States in Photonic Waveguide Arrays

Juan Kang[a,‡], Tao Liu[b,a,‡], Mou Yan[b], Dandan Yang[a], Xiongjian Huang[a], Ruishan Wei[a], Jianrong Qiu[c], Guoping Dong[a,*], Zhongmin Yang[a,*], Franco Nori[d]

[a]*State Key Laboratory of Luminescent Materials and Devices, and Guangdong Provincial Key Laboratory of Fiber Laser Materials and Applied Techniques, Guangdong Engineering Technology Research and Development Center of Special Optical Fiber Materials and Devices, School of Materials Science and Engineering, South China University of Technology, Guangzhou 510640, China*
[b]*School of Physics and Optoelectronics, South China University of Technology, Guangzhou 510640, China*
[c]*State Key Laboratory of Modern Optical Instrumentation, College of Optical Science and Engineering, Zhejiang University, Hangzhou 310027, China*
[d]*Theoretical Quantum Physics Laboratory, RIKEN Cluster for Pioneering Research, Wako-shi, Saitama 351-0198, Japan, RIKEN Center for Quantum Computing (RQC), Wako-shi, Saitama 351-0198, Japan, and Department of Physics, University of Michigan, Ann Arbor, Michigan 48109-1040, USA*

‡ These authors contributed equally.

* To whom correspondence should be addressed.
E-mail: dgp@scut.edu.cn (G.P. Dong); yangzm@scut.edu.cn (Z.M. Yang)

Dated: 2 September 2021

Recently, high-order topological insulators (HOTIs), accompanied by topologically nontrivial boundary states with codimension larger than one, have been extensively explored because of unconventional bulk-boundary correspondences. As a novel type of HOTIs, very recent works have explored the square-root HOTIs, where the topological nontrivial nature of bulk bands stems from the square of the Hamiltonian. In this paper, we experimentally demonstrate 2D square-root HOTIs in photonic waveguide arrays written in glass using femtosecond laser direct-write techniques. Edge and corner states are clearly observed through visible light spectra. The dynamical evolutions of topological boundary states are experimentally demonstrated, which further verify the existence of in-gap edge and corner states. The robustness of these edge and corner states is revealed by introducing defects and disorders into the bulk structures. Our studies provide an extended platform for realizing light manipulation and stable photonic devices.

Topological insulators (TIs) have attracted intensive research interests due to exotic electronic and optical properties as well as promising device applications[1-3]. Recently, the concept of TIs has been generalized to higher-order topological insulators (HOTIs). In contrast to first-order TIs, $n$th-order HOTIs feature[4-8] gapless states on their open boundaries with codimension $n$. For instance, a two-dimensional (2D) HOTI hosts mid-gap states on its 0D corners. The high-order topological phases have been widely investigated in both condensed matter physics[9-11] and classical waves systems, such as photonic crystals[12-20], acoustic system[21-23], and mechanical systems[24,25]. Noticeably, for photonic systems, higher-order topological states can find novel applications in, e.g., achieving 0D low-threshold and high-performance topological lasing in 2D photonic crystals[26-30], and designing topological photonic crystal fibers with multi-channel transmission capabilities based on corner modes[14,15,31].

Most recently, a novel type of HOTIs has been put forward by taking the square root of well-known models of HOTIs, named square-root HOTIs[32-34]. They are realized by adding the additional sublattices into topologically nontrivial lattices. Their Hamiltonians are obtained by applying a square-root operation to their parent Hamiltonians, which are the direct sum of the Hamiltonians of the original and inserted lattices. Topological properties of square-root Hamiltonians are inherited from their parent Hamiltonians[35-37]. In contrast to general HOTIs (e.g., kagome lattice and honeycomb lattice), hosting zero-energy corner states, a square-root HOTI hosts nonzero-energy corner states in two bandgaps[33, 34]. This provides more possibilities to manipulate photonic modes. Strikingly, when concepts of square-root HOTIs are applied to photonic systems, they have potential applications in increasing the information capacity with multi-modes in 2D photonic crystal fibers. However, the square-root HOTIs have only been experimentally demonstrated in electric circuits[33], and phononic crystals[34]. While, in photonic systems, only the 1D first-order square-root TI was recently reported in experiments[37,38]. The square-root HOTIs in two-dimensional photonic lattices remain unexplored experimentally.

In this work, we experimentally realize 2D square-root HOTIs in photonic waveguide arrays written in fused quartz glass by femtosecond (fs) laser direct-write technology[39-45]. The square-root HOTIs are constructed by inserting breathing kagome lattices into the honeycomb lattices[36]. We demonstrate the existence of high-order corner states in 2D square-root HOTIs. The dynamical features of edge and corner states are investigated by considering different lengths of waveguides. We also confirm the robustness of the edge and corner states against defects and disorder.

**Results and discussion**

**Model.** The square-root HOTI is designed by a combination of a honeycomb lattice and a breathing kagome lattice[32]. This forms a decorated honeycomb lattice, as shown in Fig. 1a. The Hamiltonian of the combined system is derived by applying the square root operator to the direct sum of two Hamiltonians of the honeycomb and breathing kagome lattices. Topological properties of the decorated honeycomb lattice are inherited from the breathing kagome lattice, i.e., hosting topologically protected second-order corner states[33,34].

We fabricate the decorated honeycomb lattice using optical waveguide arrays in the fused quartz glass by using the fs laser direct-writing technique (More details about this in the Methods and Supplementary Section 2). Each waveguide supports a single-mode guidance for the visible light. The coupling strength between two adjacent waveguides decays exponentially as their distance increases. In this case, the waveguide arrays can be described by using a tight-binding model. The diffraction equation of the light propagation along this structure reads:

$$i\partial_z \psi(z,\lambda) = H(\lambda)\psi(z,\lambda), \tag{1}$$

where $\psi(z,\lambda)$ denotes the envelope of the electric field in the waveguide with propagation distance $z$, and $\lambda$ is the wavelength of light. The effective tight-binding Hamiltonian for the decorated honeycomb lattice is given by

$$H^2 = \begin{pmatrix} H_p^{(H)} & 0 \\ 0 & H_p^{(K)} \end{pmatrix}, \tag{2}$$

where $H_p^{(H)}$ and $H_p^{(K)}$ represent the Hamiltonians of the honeycomb and breathing kagome lattices (see Fig.1a), respectively, $H_p^{(H)} = hh^\dagger$, and $H_p^{(K)} = h^\dagger h$, with

$$h = \begin{pmatrix} c_1 & c_1 & c_1 \\ c_2 & c_2 e^{ika_1} & c_2 e^{ika_2} \end{pmatrix}, \tag{3}$$

and Bloch wavevector $k$, lattice constant $a_{1,2} = \left(\pm\frac{1}{2}, \frac{\sqrt{3}}{2}\right)$ (see Fig. 1a), and coupling coefficients $c_1$ and $c_2$ between nearest-neighbor waveguides (see Fig. 1b), which depend on the wavelength $\lambda$ of light and the distance $d$ of the adjacent waveguides. The square-root Hamiltonian $H$ exhibits the topologically trivial phase for $c_1 < c_2$, and topologically nontrivial phase for $c_1 > c_2$.

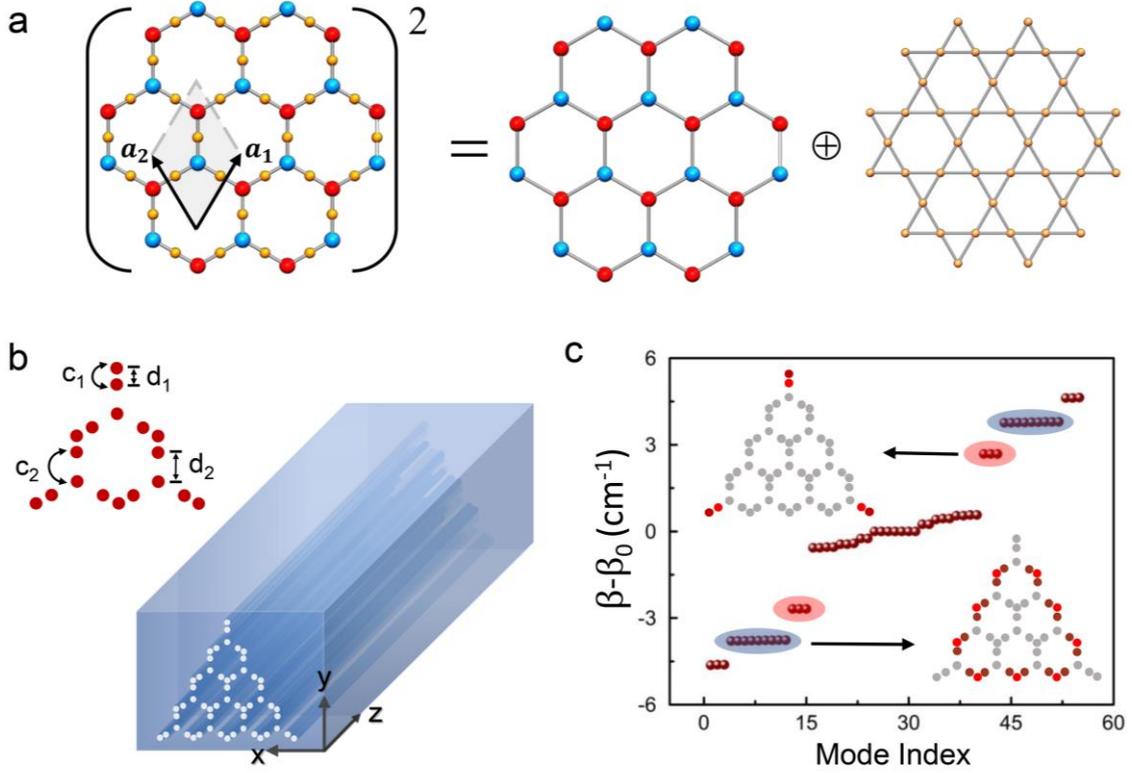

**Fig. 1 Structure and spectrum of finite decorated honeycomb lattice. a** The square of the Hamiltonian of the decorated honeycomb lattice is the direct sum of its parent Hamiltonians of honeycomb and breathing kagome lattices. Black arrows represent the lattice vectors: $a_{1,2} = \left(\pm\frac{1}{2}, \frac{\sqrt{3}}{2}\right)$. **b** Schematic showing the optical waveguide array with finite decorated honeycomb lattice fabricated in the experiment. Inset: zoom-in view of the supercell, consisting of 18 waveguides. The lattice constant $a = 33$ μm, $c_1$ and $c_2$ denote the coupling strengths defined in Eq. (3), $d_1 = 12$ μm and $d_2 = 21$ μm. **c** Calculated eigenspectrum of the finite waveguide array in **b**, corner states and edge states are marked by red and blue shadow ellipses, respectively. Mode distributions of the in-gap corner and edge states are shown in the insets.

The fabricated decorated honeycomb lattice contains 55 waveguides, as shown in Fig. 1b. The diameter of each waveguide is $8.5\,\mu m$, and the refractive index difference between the waveguide and glass matrix is estimated to be $3.3 \times 10^{-4}$, corresponding to the propagation constant $\beta_0 = 17.23$ at $\lambda = 532\,nm$. The two characteristic distances between nearest-neighbor waveguides are $d_1 = 12\,\mu m$, and $d_2 = 21\,\mu m$ (corresponding to $c_1 = 2.66\,cm^{-1}$, and $c_2 = 0.36\,cm^{-1}$, see Fig. 1b). The eigenspectra of the finite lattice are numerically calculated in Fig. 1c. There exist corner and edge modes in each bulk bandgap in topologically nontrivial phases. Moreover, corner modes are mainly localized at the two sites at each corner. While, for conventional second-order kagome lattices, corner modes are pinned at zero energy bands only in one gap, and localized at the outmost site[34]. In addition, the waveguide system will become topologically trivial once $d_1$ is interchanged with $d_2$.

**Observation of corner states and edge states.** We experimentally probe topologically protected corner and edge modes in photonic waveguide arrays, as demonstrated in Fig. 2a. A laser beam, generated from a semiconductor laser, is injected into the waveguides through the fused-tapered optical fiber. The light emerging at the output facet of the waveguide array is captured by a charge-coupled device (CCD) camera. Figure 2b shows the microscope image at the output facet ($xy$ plane). By exciting the waveguides at three different corners (see Fig. 2c-e), the light is localized at corners, indicating second-order topologically nontrivial states. The experimental results agree well with the theoretical prediction in Fig. 1c. Similarly, in order to probe edge modes, we excite the waveguide at the edge, as shown in Fig. 2f,g. The light is confined in the trimeric waveguides on the edges. Note that the light is not distributed along the edge due to its weak coupling with the neighbor trimeric waveguides. In contrast, after interchanging $d_1$ and $d_2$, the light spreads into the bulk when injected into the waveguide at either corners or edges, indicating a topologically trivial phase (see details in Supplementary Fig. S3). In addition, as demonstrated in Figs. S4 and S5, we fabricate several waveguide arrays with different coupling distances. The corner modes and edge modes are all well localized under the excitations of both blue and red lights. Note that, in photonic waveguide arrays, the coupling strength between nearest-neighbor waveguides is determined by their separations and the excitation wavelength as $c_i \approx e^{-d_i/\kappa(\lambda)}$.

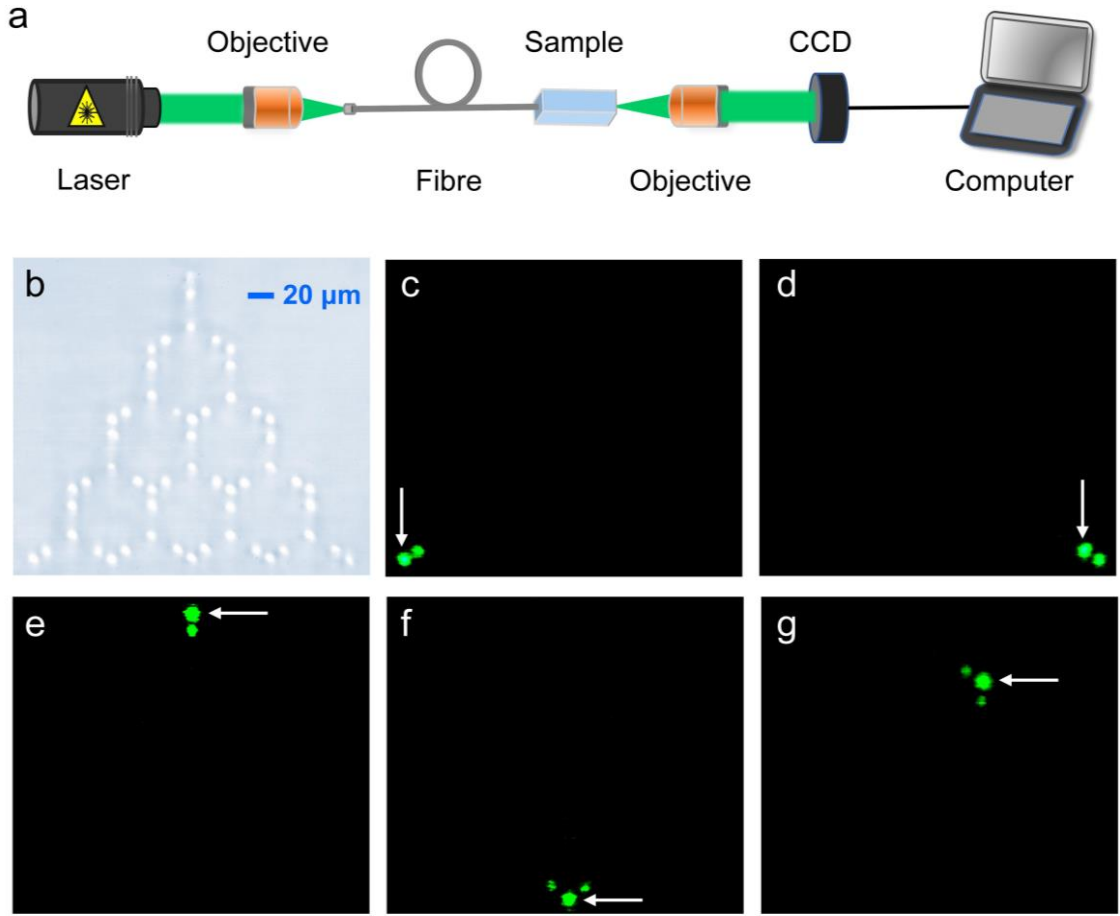

**Fig. 2 Experimental set-up and observation of topological corner (edge) states. a** Experimental setup for mode-field detections. **b** Microscope image of the decorated honeycomb lattice. **c-e** CCD camera images of light emerging at the output facet of the waveguide array with length $z = 49$ mm. Coherent light ($\lambda = 532$ nm) is injected into the waveguide at the corner (marked by a white arrow). **f, g** The same CCD camera images as **c-e**, but the light is injected into the waveguide at the edge. The images in **b-g** have the same scale.

**Dynamical evolutions.** Topological properties of the square-root HOTIs can be further explored by studying propagating dynamics, along the $z$ direction, of the corner and edge states. The evolution time $t$ of the wave function in tight-binding models can be mapped into the propagation distance $z$ of light in optical waveguide arrays. We fabricated five samples with different waveguide lengths (see Fig. 3), and probed the light distributions at the output facet. Upon exciting the first waveguide (see Figs. 3a,b and Fig. S6a) or the second waveguide (see Fig. S6) at the left corners, the light is highly localized in two waveguides at the corner. Otherwise, the light diffracts forward. When exciting the waveguide on the edge (see Figs. 3c, d), a breathing-like oscillation is observed, where light oscillates along the trimer waveguides on the edge of the waveguide array. These experimental observations are consistent with the numerical simulations by the beam-propagation method. The dynamics of the topological boundary modes further confirm the existence of stable edge and corner states protected by bulk gaps in square-root HOTIs.

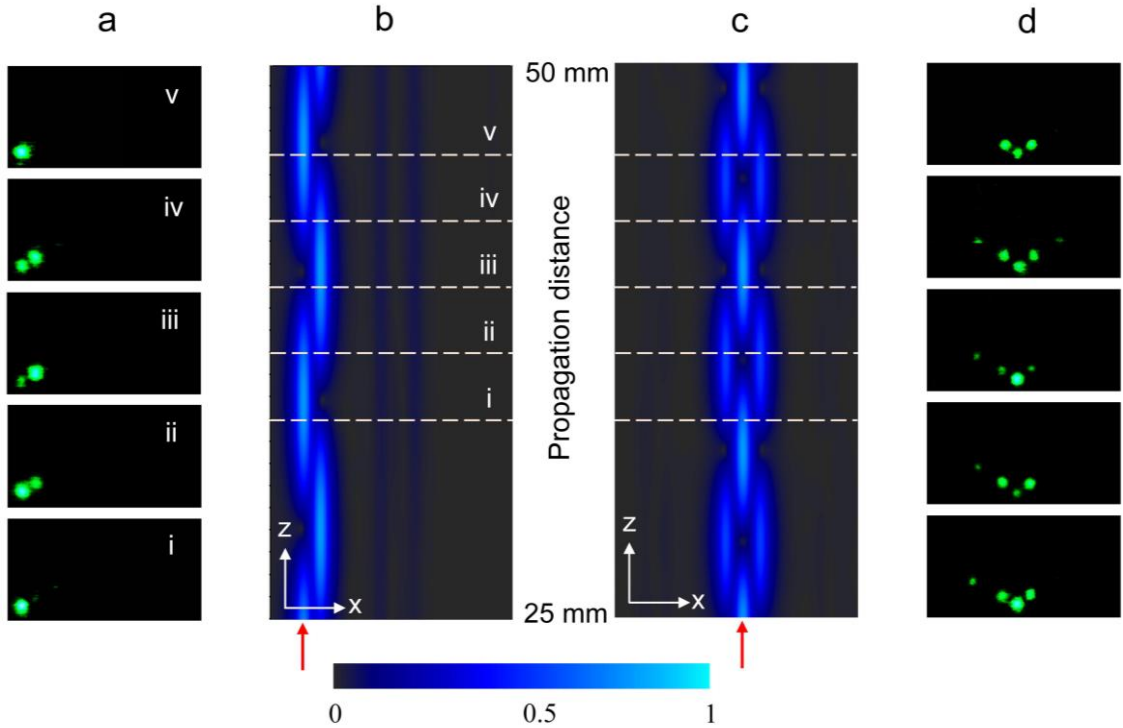

**Fig. 3 Dynamical evolutions of the corner and edge modes. a** Measured light distributions at the output facet of samples with different lengths. i-v denote waveguides with lengths $z = 34, 37, 40, 43, 46 \text{ mm}$, respectively. **b** Simulated propagation dynamics of light when exciting the first waveguides (see Fig. S6a) in **a** at the left corner. **c** Simulated propagation dynamics of the light and **d** measured light distributions when exciting the first waveguides (see Fig. S6a) at the edge. Coherent

light is injected into the input facet of the waveguide arrays, as shown by red arrows at the bottom. Note that only parts of the simulation results (propagation distance from 25 mm to 50 mm) are displayed in **b** and **c**.

**Robustness of the corner (edge) states.** To explore the robustness of the corner (edge) states against defects and disorder in the decorated honeycomb waveguide arrays, we fabricate samples by removing two waveguides in the bulk, as shown in Fig. 4a. Figures 4b-f plot the measured distributions of the output light emerging at the output facet. In spite of the defects, the corner and edge states are well localized due to topological protection. Moreover, when two waveguides on the edge are removed, the localized corner modes are also observable (see Supplementary Fig. S7). Furthermore, well-localized corner states and edge states are observed when disorders are introduced into the structure, as shown in Fig. 5 (see also Fig. S8). These results indicate that in-gap corner and edge states of square-root HOTIs are immune to defects and disorders.

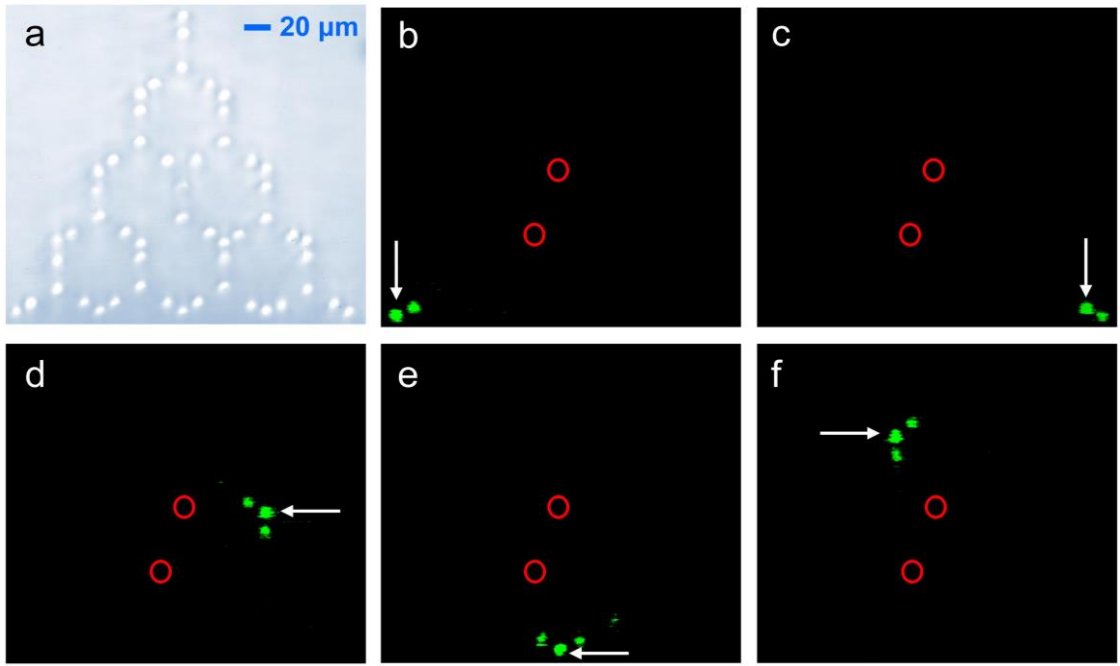

**Fig. 4 Waveguide array with bulk defects. a** Microscope image of the decorated honeycomb lattice with two missing waveguides in the bulk. **b-c** CCD camera images of light emerging at the output facet of the waveguide array with bulk defects ($z =$

49 mm). Coherent light is injected into the waveguide at the corner (white arrow). Missing waveguides are marked by red circles. **d-f** Same CCD camera images as **b-c**, but the light is injected into the waveguides on the edge.

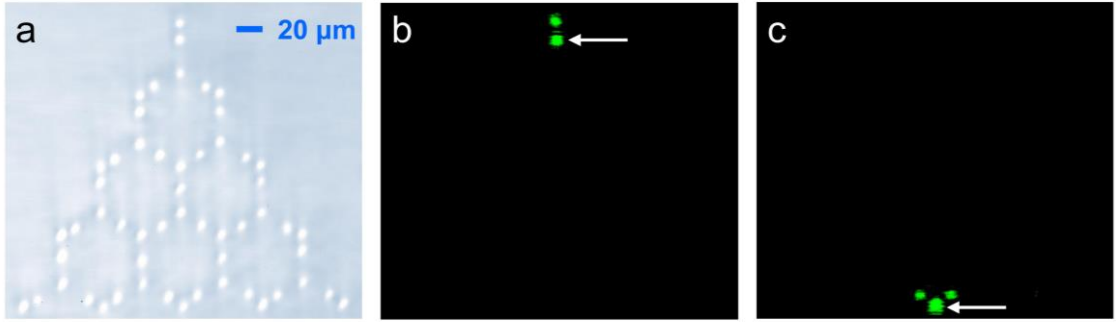

**Fig. 5 Waveguide array with disorder. a** Microscope image of the decorated honeycomb lattice with disorder in the bulk. **b-c** CCD camera images of light emerging at the output facet of the waveguide array with the disorder ($z = 49$ mm). Coherent light is injected into the waveguides at the corner for **b,** and the edge for **c** (indicated by a white arrow), respectively.

**Conclusions**

In summary, we experimentally observe higher-order topologically protected corner states in square-root HOITs based on optical waveguide arrays. We fabricated decorated honeycomb lattice structures, consisting of breathing kagome and honeycomb lattices, written in glass using femtosecond laser direct-write techniques. We observe corner states localized at the two corner waveguides of the corners are observed. The dynamical evolutions of corner states and edge states are experimentally demonstrated. These topologically protected modes are shown to be insensitive to defects and disorder. We have, for the first time, provided an optical realization of novel higher-order topological phases. Our simple and stable photonic structures can drastically enhance the prospects of implementing these ideas in photonic technological devices, e.g., improving nonlinearity and exhibiting robustness for controlling light propagation in optical fibers[46-48].

**Methods**

**Sample fabrication.** Waveguide arrays were fabricated with $50\,\mathrm{mm}$ commercial quartz glasses by using fs laser processing. This system consists of a regeneratively amplified Ti: sapphire fs laser, a Nikon microscope (Eclipse 80i) equipped with a CCD camera, a computer-controlled $3\mathrm{D}\,xyz$ translation stage, and several optical elements. The fs laser emits $1\,\mathrm{kHz}$, $130\,\mathrm{fs}$ pulses with central wavelength of $800\,\mathrm{nm}$. To write the waveguide arrays, laser pulses with the energy of $2.5\,\mathrm{mW}$ were focused onto the sample surface (with diameters of $110-314\,\mathrm{\mu m}$) using a $20\times(NA\,0.45)$ objective. The glass sample was placed on the $3\mathrm{D}\,xyz$ translation stage, and then translated at a speed of $300\,\mathrm{\mu m/s}$. An adjustable slit, oriented parallel to the laser writing direction, is inserted in front of the objective to shape the laser beam and fabricate the near-circular cross-section waveguide, (see details in Supplementary Section 2). The fabricated waveguides support a single mode guide over the entire visible wavelength range. After the waveguide arrays were fabricated, the two lateral faces of the sample were carefully polished. Finally, $49\,\mathrm{mm}$ waveguide arrays samples were used for the observation of topological phenomena.

**Acknowledgements**

This work was financially supported by the Key R&D Program of Guangzhou (202007020003), National Natural Science Foundation of China (Grant Nos. 5200020611, 62075063, 51772101, 51872095), the fellowship of China Postdoctoral

Science Foundation (2020M672621, 2021M691054), Local Innovative and Research Teams Project of Guangdong Pearl River Talents Program (2017BT01X137). T.L. acknowledges the support from the Startup Grant of South China University of Technology (Grant No. 20210012). F.N. is supported in part by: Nippon Telegraph and Telephone Corporation (NTT) Research, the Japan Science and Technology Agency (JST) [via the Moonshot RD Grant Number JPMJMS2061, and the Centers of Research Excellence in Science and Technology (CREST) Grant No. JPMJCR1676], the Japan Society for the Promotion of Science (JSPS) [via the Grants-in-Aid for Scientific Research (KAKENHI) Grant No. JP20H00134 and the JSPS{RFBR Grant No. JPJSBP120194828], the Army Research Officece (ARO) (Grant No. W911NF-18-1-0358), the Asian Officece of Aerospace Research and Development (AOARD) (via Grant No. FA2386-20-1-4069), and the Foundational Questions Institute Fund (FQXi) via Grant No. FQXi-IAF19-06.

## Author contributions

All authors contributed extensively to the work presented in this paper.

## Conflict of Interest

The authors declare no conflict of interest.